\documentclass[aip,reprint,superscriptaddress,nofootinbib]{revtex4-1} 
\usepackage[english]{babel}

\usepackage{amsmath, amssymb}
\usepackage{mathtools}
\usepackage{siunitx}
\usepackage{environ}
\usepackage{mathrsfs}
\usepackage{braket}
\usepackage{bm}
\DeclareMathAlphabet{\mathscrlower}{OT1}{pzc}{m}{it} 
\usepackage{prettyref}
\usepackage{graphicx}
\usepackage{booktabs}
\usepackage{array}
\usepackage{multirow}
\usepackage{dcolumn}
\usepackage{threeparttable}
\usepackage{threeparttablex}
\usepackage{placeins}
\usepackage[stable]{footmisc}
\usepackage{mhchem}
\usepackage[colorlinks]{hyperref}

\newcommand{\pauli}{\boldsymbol{\sigma}}
\newcommand{\Pauli}{\boldsymbol{\sigma}}
\newcommand{\diraccontra}[1]{\boldsymbol{\gamma}^{#1}}

\newcommand{\diraca}{\vec{\boldsymbol{\alpha}}}
\newcommand{\diracg}{\vec{\boldsymbol{\gamma}}}
\newcommand{\diracb}{\boldsymbol{\beta}}
\newcommand{\unity}{\bm{1}_{2\times 2}}

\newcommand{\pos}{\vec{r}}

\newcommand{\momop}{\hat{\vec{p}}}

\newcommand{\spinmom}{\vec{\Pauli}\cdot\momop}

\newcommand{\Sum}[2]{\sum\limits_{#1}^{#2}}

\newcommand{\parantheses}[1]{\left(#1\right)}
\newcommand{\brackets}[1]{\left[#1\right]}
\newcommand{\braces}[1]{\left\{ #1\right\}}
\let\nablatmp\nabla
\renewcommand{\nabla}{\vec{\nablatmp}}
\DeclarePairedDelimiter\abs{\lvert}{\rvert}
\makeatletter
\let\oldabs\abs
\def\abs{\@ifstar{\oldabs}{\oldabs*}}

\newcommand{\Op}[1]{\hat{#1}}

\newcommand{\partiell}[2]{\frac{\partial #1}{\partial #2}}
\AtBeginDocument{
\heavyrulewidth=.08em
\lightrulewidth=.05em
\cmidrulewidth=.03em
\belowrulesep=.65ex
\belowbottomsep=0pt
\aboverulesep=.4ex
\abovetopsep=0pt
\cmidrulesep=\doublerulesep
\cmidrulekern=.5em
\defaultaddspace=.5em
}

\begin{document}
\title{Toolbox approach for quasi-relativistic calculation of molecular 
properties for precision tests of fundamental physics}
\date{\today}
\author{Konstantin Gaul}
\affiliation{Fachbereich Chemie, Philipps-Universit\"{a}t Marburg,
Hans-Meerwein-Stra\ss{}e 4, 35032 Marburg, Germany}
\author{Robert Berger}
\affiliation{Fachbereich Chemie, Philipps-Universit\"{a}t Marburg,
Hans-Meerwein-Stra\ss{}e 4, 35032 Marburg, Germany}
\begin{abstract}
A generally applicable approach for the calculation of relativistic
one-electron properties with two-component wave functions is presented. The
formalism is explicitly evaluated for the example of quasi-relativistic
wavefunctions obtained within the zeroth order regular approximation
(ZORA). The wide applicability of the scheme is demonstrated for the
calculation of parity ($\mathcal{P}$) and time-reversal ($\mathcal{T}$)
symmetry violating properties, which are important for searches of physics
beyond the standard model of particle physics. The quality of the ZORA
results is shown exemplarily for the molecules RaF and TlF by comparison
to data from four-component calculations as far as available. Finally, the 
applicability of RaF in experiments that search for $\mathcal{P,T}$-violation 
not only in the electronic but also in quark sector is demonstrated.
\end{abstract}
\maketitle
\section{Introduction}
In order to explore the boundaries of the standard model of particle
physics and of general relativity, a wealth of experiments was proposed in
the last decades (see e.g. Ref. \onlinecite{demille:2017}) aiming for the
detection of new physics. Molecular systems have gained meanwhile
increasing importance for this research direction, which has recently been
reviewed for instance in Ref. \onlinecite{safranova:2018}. Modern
experiments take advantage of the rich, but also highly complex vibronic
internal structure of molecules, which is why theory plays a crucial role
for design and interpretation of established and future experiments.

From this search for new physics emerges also a large number of new
properties of interest, of which many violate symmetries that are either
conserved or only very weakly broken in established physics (see e.g.  Ref.
\onlinecite{khriplovich:1997}). Several of the proposed phenomena are
considered to be favourably enhanced in heavy-elemental molecules due to
relativistic effects and therefore demand on the theory side a
corresponding description of the electronic wave function, including
spin-orbit coupling.

An accurate treatment of spin-orbit coupling in heavy elements requires at
least a two-component wave function, which is nowadays available in many
quantum chemistry software packages. However, naturally relativistic
properties are typically described in a four-component framework.
Therefore, all operators have to be transformed from four-component to
two-component pictures in order to be consistent with quasi-relativistic
wave function formulations. As most programs do not include direct support
for four-component treatments, this requires derivation of analytical
expressions of two-component operators and commonly also a new, often tedious 
and error-prone implementation for every property.

In this paper we develop a general formalism for flexible calculation
of arbitrary relativistic one-electron properties in a two-component
framework. We introduce our formalism in terms of two-component
one-electron density functions and we provide a general formulation of
relativistic one-electron properties for the case of zeroth order
regular approximation (ZORA) wave functions. Our generally applicable
scheme is demonstrated on the example of different sources of
simultaneous parity ($\mathcal{P}$) and time-reversal ($\mathcal{T}$)
symmetry violation in the diatomic molecules RaF and TlF, which are
promising candidates for a first measurement of
$\mathcal{P,T}$-violation beyond the Standard
Model.\cite{hinds:1980,hunter:2012,Isaev:2013,norrgard:2017} 
In future applications the concepts and implementation derived in the
following can also be used to study conventional nuclear magnetic
resonance (NMR) shielding constants and parity violating
shifts in chiral molecules\cite{nahrwold:09,nahrwold:2014} and a
wealth of further properties.

\section{Theory\label{theory}}
\subsection{Relativistic one-particle wave functions and
operators}
Relativistic electronic one-particle wave 
functions have a bi-spinor structure, which can be decomposed as
\begin{equation}
\psi
=
\begin{pmatrix}
\psi^\text{L}\\
\psi^\text{S}
\end{pmatrix}
=\begin{pmatrix}
\psi^\text{L}_\alpha\\
\psi^\text{L}_\beta\\
\psi^\text{S}_\alpha\\
\psi^\text{S}_\beta
\end{pmatrix}.
\end{equation}
Here $\psi^\text{L}$ and $\psi^\text{S}$ are the spinors of the large
or upper component and the small or lower component of
the Dirac one-particle wave function $\psi$, respectively. $\psi_\alpha$ 
and $\psi_\beta$ are spin orbitals with electron spin up and down,
respectively.
A similar structure is also found for one-particle operators in four component
theory. Operators can be decomposed in a $2\times2$-block structure:
\begin{equation}
\Op{\mathbf{O}}=\begin{pmatrix}
\hat{\mathbf{O}}^{\text{LL}}_{2\times2}&&\hat{\mathbf{O}}^{\text{LS}}_{2\times2}\\
\hat{\mathbf{O}}^{\text{SL}}_{2\times2}&&\hat{\mathbf{O}}^{\text{SS}}_{2\times2}
\end{pmatrix}
\end{equation} 
Thus, we can deduce four kinds of one-particle densities, which appear as
contribution to all expectation values of custom one-electron operators in
one-particle four-component theory. These are contributions from large
component-large component (LL), large component-small component (LS), small
component-large component (SL) and small component-small component
(SS) matrix elements. This formulation allows a decomposition of
expectation values of four-component one-electron operators in a sum of modified
relativistic one-electron density functions
$\Omega^{IJ}(\vec{x})=\parantheses{\psi^I(\vec{x})}^\dagger\hat{\mathbf{O}}^{IJ}_{2\times2}\psi^J(\vec{x})$,
where $IJ$ can be LL, LS, SL or SS:
\begin{equation}
\Braket{\hat{\mathbf{O}}}=\int\mathrm{d}^4\vec{x}~\brackets{
 \Omega^\mathrm{LL}(\vec{x})
+\Omega^\mathrm{SL}(\vec{x})
+\Omega^\mathrm{LS}(\vec{x})
+\Omega^\mathrm{SS}(\vec{x})
}.
\end{equation}
In the following all $\Omega^{IJ}$ are reformulated in terms of
two-component density functions exploiting the Lorentz symmetry of
relativistic operators. Finally, this reformulation will allow a
generally applicable procedure for computation of $\Omega^{IJ}$ within
an approximate two-component theory.
\subsection{Lorentz group and relativistic density functions}
The explicit representation of four-component one-electron operators
can be reduced to a linear-combination of a special unitary basis of
matrices in four-dimensional space that corresponds to the Lorentz group
$SU(2)\times SU(2)$, which represents the $2\times2$-block structure
of the Dirac equation. This basis is formed by the 16 (or rather 32,
allowing an imaginary phase of $\imath^k$ with the imaginary unit
$\imath=\sqrt{-1}$ and $k\in\mathbb{N}_0$) Dirac matrices:\\
\begin{equation}
\boldsymbol{\Gamma}^{i,j,k}=\imath^k\pauli^i\otimes\pauli^j;~i,j=0,1,2,3~\wedge~k\in\mathbb{N}_0,
\end{equation}
where $\pauli^i$ for $i=1,2,3$ are the Pauli spin matrices and $\pauli^0=\unity$
is the two dimensional identity matrix, which build the basis of
$SU(2)$, and $\otimes$ is the Kronecker product. We include an
imaginary phase $\imath^k$ in order to enable time-reversal symmetry
violation in our formalism, as well.

The usual notation for theses matrices is 
\begin{subequations}
\begin{align}
\boldsymbol{\Gamma}^{0,0,k}&=&  \imath^k\pauli^0\otimes\pauli^0  &=&\imath^k{\bf 1}_{4\times4}                    \\
\boldsymbol{\Gamma}^{1,0,k}&=&  \imath^k\pauli^1\otimes\pauli^0  &=&\imath^k\diraccontra{5}                       \\
\boldsymbol{\Gamma}^{2,0,k}&=&
\imath^{k-1}\imath\pauli^2\otimes\pauli^0&=&\imath^{k-1}\diraccontra{0}\diraccontra{5}        \\
\boldsymbol{\Gamma}^{3,0,k}&=&  \imath^k\pauli^3\otimes\pauli^0  &=&\imath^k\diraccontra{0}=\imath^k\diracb           \\
\vec{\boldsymbol{\Gamma}}^{0,(1,2,3),k}&=&\imath^k\pauli^0\otimes\vec{\pauli}  &=&\imath^k\vec{\boldsymbol{\Sigma}}               \\
\vec{\boldsymbol{\Gamma}}^{1,(1,2,3),k}&=&\imath^k\pauli^1\otimes\vec{\pauli}  &=&\imath^k\diraccontra{0}\diracg=\imath^k\diraca          \\
\vec{\boldsymbol{\Gamma}}^{2,(1,2,3),k}&=&\imath^{k-1}\imath\pauli^2\otimes\vec{\pauli}&=&\imath^{k-1}\diracg                                 \\
\vec{\boldsymbol{\Gamma}}^{3,(1,2,3),k}&=&\imath^k\pauli^3\otimes\vec{\pauli}
&=&\imath^k\diraccontra{0}\vec{\boldsymbol{\Sigma}} ~.
\label{eq: dirac}
\end{align}
\end{subequations}
Employing this structure we can define 32 Dirac one-electron density
functions of electron 1 which are build as
\begin{multline}
\Gamma^{i,j,k}(\pos_1,\pos_1')
\\
=\hat{\mathcal{D}}\brackets{
\parantheses{\Psi(\vec{X})}^\dagger
\vec{\boldsymbol{\Gamma}}_1^{i,j,k}
\otimes\boldsymbol{\Gamma}^{0,0,0}_2\otimes\cdots\otimes\boldsymbol{\Gamma}^{0,0,0}_{N_\mathrm{elec}}
\Psi(\vec{X}')}.
\end{multline}
We have introduced in this equation the many-particle wave function $\Psi$, 
which depends on the set $\vec{X}=\braces{\vec{x}_1,\dots,\vec{x}_N}$ of
$N_\mathrm{elec}$ combined spin and spatial coordinates, with
$\vec{x}_i=\braces{\sigma_i,\pos_i}$ for each electron $i$, and the
one-electron density projection operator 
\begin{equation}
\hat{\mathcal{D}}=N_\mathrm{elec}\mathfrak{Re}
\int\limits_{\sigma_1=\sigma_1'}\mathrm{d}\sigma_1\idotsint\limits_{\vec{x}_i=\vec{x}'_i;~i=1,\dots,N}
\mathrm{d}^4\vec{x}_2\cdots\mathrm{d}^{4}\vec{x}_N,
\end{equation}
which integrates out all coordinates except one spatial coordinate
$\vec{r}_1$. Coordinates with a prime are introduced to formally allow
a discrimination between the left and right function that contributes
to the density function, in order to enable the definition of
operators acting on the left or right function, respectively. 

Employing the $2\times2$-block structure of the one-particle Dirac
equation we can decompose the Dirac one-electron density functions further by
introducing four kinds of two-component densities for each of the four
four-component densities $\rho^{IJ,j,k}$. The two-component density functions can be
classified as i) number density functions $\rho^{IJ,0,2k}$, ii) number current
density functions $\rho^{IJ,0,2k+1}$, iii) spin density functions
$\vec{\rho}^{IJ,(1,2,3),2k}$ and iv) spin current density functions
$\vec{\rho}^{IJ,(1,2,3),2k+1}$. Thus there are in total $4\times4=16$
two-component one-electron density functions, which can be combined linearly to
give the 32 Dirac one-electron density functions: 
\begin{subequations}
\begin{align}
\Gamma^{0,j,k}(\pos,\pos')&=\rho^{\mathrm{LL},j,k}(\pos,\pos')+\rho^{\mathrm{SS},j,k}(\pos,\pos')\\
\Gamma^{1,j,k}(\pos,\pos')&=\rho^{\mathrm{LS},j,k}(\pos,\pos')+\rho^{\mathrm{SL},j,k}(\pos,\pos')\\
\Gamma^{2,j,k}(\pos,\pos')&=\rho^{\mathrm{LS},j,k}(\pos,\pos')-\rho^{\mathrm{SL},j,k}(\pos,\pos')\\
\Gamma^{3,j,k}(\pos,\pos')&=\rho^{\mathrm{LL},j,k}(\pos,\pos')-\rho^{\mathrm{SS},j,k}(\pos,\pos')~.
\end{align}
\label{eq: diracdensity}
\end{subequations}
We introduce the spin tensor function in correspondence to the four
Pauli matrices with a complex phase:
\begin{equation}
\boldsymbol{\varsigma}(j,k)=
\imath^k\pauli^j~.
\end{equation}
Using this spin tensor function the two-component one-electron density
functions of electron 1 are build with the spinors of the Dirac
one-particle wave function containing either only the large component
for the first electron $\Psi^\mathrm{L}$ or only small component for
the first electron $\Psi^\mathrm{S}$ as
\begin{multline}
\rho^{IJ,j,k}(\vec{r}_1,\vec{r}'_1) 
\\
=
\hat{\mathcal{D}}\brackets{
\parantheses{\Psi^I(\vec{X})}^\dagger
\boldsymbol{\varsigma}_1(j,k)
\otimes\pauli^0_2\otimes\cdots\otimes\pauli^0_{N_\mathrm{elec}}
\Psi^J(\vec{X}')}.
\label{eq: densityfunctions}
\end{multline}

\subsection{Quasi-relativistic approximation and ZORA}
In order to be able to calculate these relativistic density functions
in two-component theories we need to approximate the small component
wave function. The exact transformation of the large-component
one-particle wave function into the small-component one-particle wave
function is given by:
\begin{multline}
\psi^\mathrm{S}(\vec{x})=
\underbrace{
c\parantheses{2m_\mathrm{e}c^2\unity-\Op{\mathbf{V}}^{\mathrm{SS}}_\mathrm{diag}(\vec{x})+\epsilon\unity}^{-1}
}_{=c\boldsymbol{\omega}(\vec{x})}\\
\times 
\underbrace{\parantheses{\vec{\pauli}\cdot\momop+\Op{\mathbf{V}}^{\mathrm{SL}}_\mathrm{off}(\vec{x})}}_{\boldsymbol{\varpi}(\vec{x})}
\psi^\mathrm{L}(\vec{x})
.
\end{multline}
Here $\epsilon$ is the electronic energy of the single particle state,
$\momop=-\imath\hbar\nabla$ is the one-electron linear momentum
operator in position space with $\hbar=\frac{h}{2\pi}$ being the reduced 
Planck's constant, $\Op{\mathbf{V}}^{\mathrm{SS}}_\mathrm{diag}$ is the
SS-block of the potential effective one-electron operator appearing on
the diagonal of the Hamiltonian,
$\Op{\mathbf{V}}^{\mathrm{SL}}_\mathrm{off}$ is the SL-block of a
potential effective one-electron operator appearing on the
off-diagonal of the Hamiltonian, as e.g. a vector potential or the
Breit operator, both as effective one-electron operators.
$m_\mathrm{e}$ is the electron mass and $c$ is the 
speed of light in vacuum.

For any two-component method we can replace $\psi^\mathrm{L}$
approximately by the wave-function optimized within this specific
method and use the corresponding approximate transformation matrix
$\boldsymbol{\omega}$ to construct an approximate $\psi^\mathrm{S}$.

As $\Op{\mathbf{V}}^{\mathrm{SL}}_\mathrm{off}(\vec{x})$ is an
effective one-electron operator, the modified momentum matrix
$\boldsymbol{\varpi}$ can be decomposed into a linear combination of spin
tensor functions:
\begin{equation}
\boldsymbol{\varpi}(\vec{x})=\Sum{j=0}{3}\Sum{k=0}{1}
\boldsymbol{\varsigma}(j,k)
\hat{\pi}^{j,k}(\vec{x})
\end{equation}
with the modified one-electron momentum operators $\hat{\vec{\pi}}^{j,k}$.
Neglecting all off-diagonal potentials, $\boldsymbol{\varpi}$ simplifies to
\begin{equation}
\boldsymbol{\varpi}\approx\spinmom=-\hbar\imath\vec{\pauli}\cdot\nabla,
\end{equation}
so that $\hat{\vec{\pi}}^{(1,2,3),1}=-\hbar\nabla$. For all
$j=0$ or $k=0$, we have $\hat{\pi}^{j,k}_i=0$.

In the present paper we implemented the scheme within zeroth order
regular approximation (ZORA) without potentials appearing on
the off-diagonal:
\begin{align}
\psi^\text{L}(\vec{x})&\approx\psi^\text{ZORA}(\vec{x})\\
\psi^\text{S}(\vec{x})&\approx\underbrace{\frac{c}{2m_\mathrm{e}c^2-\tilde{V}(\pos)}}_{=c\omega^\text{ZORA}(\pos)}\vec{\pauli}\cdot\momop\psi^\text{ZORA}_i(\vec{x}),
\end{align}
where we assume an unperturbed scalar transformation factor $\omega$,
which depends only on a model-potential $\tilde{V}$ as introduced by
van W\"ullen.\cite{wullen:1998} The model-potential is an
effective one-electron operator that does not depend on the molecular
wave function and is designed to alleviate the gauge variance of ZORA. All 
perturbations to the diagonal potential
$\Op{\mathbf{V}}_\mathrm{diag}$ are neglected in $\tilde{V}$.

Corrections stemming from the presence of vector potentials appear as
additional terms and can be included afterwards by calculation of matrix
elements of the corresponding operators, which is
straightforward in the present approach.

In the mean-field approximation the ZORA $N_\mathrm{elec}$-electron wave
function is represented as $N_\mathrm{elec}$ Slater determinant of 
one-electron wave functions $\psi_i^\text{ZORA}$, which are composed of
$N_\mathrm{basis}$ real spatial basis functions $\chi_\mu$ and complex two
component coefficients $\vec{C}_{i\mu}$:
\begin{equation}
\Psi^\text{ZORA}=\abs{\psi_1^\text{ZORA}\cdots\psi_N^\text{ZORA}}
\end{equation}
with
\begin{equation}
\psi_i^\text{ZORA}=\Sum{\mu=1}{N_\text{basis}}\vec{C}_{i\mu}\chi_\mu,
\end{equation}
where
\begin{equation}
\vec{C}_{i\mu}=\begin{pmatrix}C^\alpha_{i\mu}\\C^\beta_{i\mu}\end{pmatrix}.
\end{equation}
Only the coefficients depend on the spin component, so that 
spin can in this form easily be integrated out. 
Furthermore, in this approach we can define for each molecular orbital
(MO) $i$ a MO-density function $\rho^{IJ,j,k}_i$. The total
ZORA-density function is a sum of all $N_\text{orb}$ MO-density
functions weighted by their occupation number $n_i$, which is 0 for
unoccupied or 1 for occupied orbitals:
\begin{equation}
\rho^{IJ,j,k}(\pos,\pos')=\Sum{i=1}{N_\text{orb}}n_i\rho^{IJ,j,k}_i(\pos,\pos').
\end{equation}
Within ZORA the one-electron MO-density functions
$\rho^{IJ,j,k}_i$
are evaluated explicitly as
\begin{widetext}
\begin{subequations}
\begin{align}
\rho^{\mathrm{LL},j,k}_i(\pos,\pos')=&
\Sum{\mu\nu}{N_\text{basis}}
\underbrace{\mathfrak{Re}\braces{\vec{C}_{i\mu}^\dagger\vec{\boldsymbol{\varsigma}}(j,k)\vec{C}_{i\nu}}}
_{\mathcal{D}_{i\mu\nu}^{j,k,(0,0)}}
~\cdot~
\underbrace{\parantheses{\chi_\mu\chi'_\nu}}
_{\mathcal{I}^{(0,0)}_{\mu\nu}}
\label{eq: 1eMODF1}
\\
\rho^{\mathrm{LS},j,k}_i(\pos,\pos')=&
\Sum{\mu\nu}{N_\text{basis}}
\Sum{l=1}{3}
\underbrace{\mathfrak{Re}\braces{\vec{C}_{i\mu}^\dagger
\boldsymbol{\varsigma}(j,k)
\parantheses{-\imath\pauli^l\vec{C}_{i\nu}}}}
_{\mathcal{D}_{i\mu\nu}^{j,k,(0,l)}}
~\cdot~
\underbrace{\parantheses{\hbar c\chi_\mu\omega'(\vec{\partial}'_\nu)_l}}
_{\mathcal{I}^{(0,l)}_{\mu\nu}}
\label{eq: 1eMODF2}
\\
\rho^{\mathrm{SL},j,k}_i(\pos,\pos')=&
\Sum{\mu\nu}{N_\text{basis}}
\Sum{l=1}{3}
\underbrace{\mathfrak{Re}\braces{\parantheses{\vec{C}_{i\mu}^\dagger\imath\pauli^l}
\boldsymbol{\varsigma}(j,k)
\vec{C}_{i\nu}}}
_{\mathcal{D}_{i\mu\nu}^{j,k,(l,0)}}
~\cdot~
\underbrace{\parantheses{\hbar c(\vec{\partial}_\mu)_l\omega\chi'_\nu}}
_{\mathcal{I}^{(l,0)}_{\mu\nu}}
\label{eq: 1eMODF3}
\\
\rho^{\mathrm{SS},j,k}_i(\pos,\pos')=&
\Sum{\mu\nu}{N_\text{basis}}
\Sum{l=1}{3}
\Sum{m=1}{3}
\underbrace{\mathfrak{Re}\braces{\parantheses{\vec{C}_{i\mu}^\dagger\imath\pauli^l}\boldsymbol{\varsigma}(j,k)\parantheses{-\imath\pauli^m\vec{C}_{i\nu}}}}
_{\mathcal{D}_{i\mu\nu}^{j,k,(l,m)}}
~\cdot~
\underbrace{\parantheses{\hbar^2c^2(\vec{\partial}_\mu)_l\omega\omega'(\vec{\partial}'_\nu)_m}}
_{\mathcal{I}^{(l,m)}_{\mu\nu}}
.
\label{eq: 1eMODF4}
\end{align}
\label{eq: 1eMODF}
\end{subequations}
\end{widetext}
Here we used the short-hand notations: $\chi_\mu(\pos)=\chi_\mu$,
$\chi_\mu(\pos')=\chi'_\mu$, 
$\nabla\chi_\mu=\vec{\partial}_\mu$,
$\nabla'\chi'_\mu=\vec{\partial}'_\mu$,
$\omega^\text{ZORA}(\pos)=\omega$ and
$\omega^\text{ZORA}(\pos')=\omega'$.
Generalization of eqs. \prettyref{eq: 1eMODF1} to \prettyref{eq:
1eMODF4} in terms of
$\mathcal{D}$, $\mathcal{I}$ gives 
\begin{equation}
\rho^{IJ,j,k}_i(\pos,\pos')=\Sum{\mu\nu}{N_\mathrm{basis}}\Sum{
l\in M(I)}{}
\Sum{
m\in M(J)}{}
\mathcal{D}^{j,k,(l,m)}_{i\mu\nu}\mathcal{I}^{(l,m)}_{\mu\nu}(\pos,\pos'),
\label{eq: 1eMODF_split}
\end{equation}
where $M(I)= \begin{cases}
\braces{1,2,3}&\text{if}~I=\mathrm{S}\\
\braces{0}    &\text{if}~I=\mathrm{L}
\end{cases}$ is a mapping between spin indices $l,m$ of
$\boldsymbol{\mathcal{D}}^{j,k,(l,m)}_i$ and
$\boldsymbol{\mathcal{I}}^{(l,m)}$ and the relativistic type of the
two-component density function $I,J$ and
\begin{align}
\nonumber
&\mathcal{D}^{j,k,(l,m)}_{i\mu\nu}
\\
&=
\mathfrak{Re}\braces{
  \parantheses{\vec{C}_{i\mu}}^\dagger
  \boldsymbol{\varsigma}(l,1-\delta_{l0})
  \boldsymbol{\varsigma}(j,k)
  \boldsymbol{\varsigma}(m,3-3\delta_{m0})
  \vec{C}_{i\nu}
}
\label{eq: twocompDF4c}
\\
&\mathcal{I}^{(l,m)}_{\mu\nu}(\pos,\pos')
=
  (\hbar c)^{2-\delta_{l0}-\delta_{m0}}
  \partial^{(l)}_\mu 
  \omega^{1-\delta_{l0}}
  \omega'^{1-\delta_{m0}}
  \partial'^{(m)}_\nu\,,
\label{eq: twocompAO4c}
\end{align}
with $\partial^{(i)}_\mu =\begin{cases}
\partiell{\chi_\mu}{(\vec{r})_i}&\text{if}~i=1,2,3\\
\chi_\mu                          &\text{if}~i=0\end{cases}
$ and analogously
with prime.

Here the density matrices
$\boldsymbol{\mathcal{D}}^{j,k,(l,m)}=\Sum{i=1}{N_\mathrm{occ}}\boldsymbol{\mathcal{D}}^{j,k,(l,m)}_i$
can be constructed from the four (eight when separating real and
imaginary parts) two-component density matrices $\mathbf{D}^{(k)}$,
which are defined in terms of Pauli matrices as
\begin{equation}
D^{(k)}_{\mu\nu}=\Sum{i=1}{N_\text{occ}}\underbrace{\vec{C}^\dagger_{i\mu}\pauli^k\vec{C}_{i\nu}}_{D^{(k)}_{i\mu\nu}}.
\label{eq: density_matrix}
\end{equation}
The explicit expressions for $\boldsymbol{\mathcal{D}}^{j,k,(k,l)}$ in
terms of $\mathbf{D}^{(k)}$ can be evaluated following eq.
\prettyref{eq: twocompDF4c} and using
commutator relations of Pauli matrices.
\subsection{Generic tensor function formulation of molecular
properties}
With the two-component formulation of relativistic density functions
introduced in the previous section we can formulate a generic
one-electron tensor function which can directly be connected to an
arbitrary one-electron molecular property. For this purpose we introduce,
besides the relativistic $\Gamma^{i,j,k}$ or quasi-relativistic
one-electron density function $\rho^{IJ,j,k}$,
a general differential tensor operator
$\hat{\boldsymbol{\partial}}(\pos'\vee\pos)$, which acts on $\pos'$ or
$\pos$. Within this flexible formulation we do not use
turn-over rule, which can complicate calculations as pointed out in
Ref. \onlinecite{kutzelnigg:1988}. Furthermore a general tensor operator
$\hat{\bm{\mathsf{t}}}(\pos)$ and a
general scalar operator $\hat{s}$ are defined. All these operators may be 
defined arbitrarily. In the following section, some explicit realisations for 
these operators will be discussed. A generic one-electron tensor function
can now be written as
\begin{equation}
\boldsymbol{\Omega}^{i,j,k}(\pos)=
\hat{s}\hat{\bm{\mathsf{t}}}\circ\left.\hat{\boldsymbol{\partial}}(\pos'\vee\pos)\circ
\Gamma^{i,j,k}(\pos,\pos')\right|_{\pos'=\pos},
\end{equation}
or for two-component density functions as
\begin{equation}
\boldsymbol{\Omega}^{IJ,j,k}(\pos)=
\hat{s}\hat{\bm{\mathsf{t}}}\circ\left.\hat{\boldsymbol{\partial}}(\pos'\vee\pos)\circ
\rho^{IJ,j,k}(\pos,\pos')\right|_{\pos'=\pos},
\end{equation}
where $\circ=\otimes~,\cdot,~\times$, with $\otimes$ being the outer
product, $\cdot$ being the inner product and $\times$ being the cross
product. The latter is defined only within $\mathbb{R}^3$. Here $j$
can be $0$ or $(1,2,3)$. In the latter case $\Gamma^{i,j,k}$ is a
three dimensional vector.
This formulation allows the flexible construction of one-electron
operators that can correspond to molecular properties. The expectation
value is received by integration over space:
\begin{equation}
\Braket{\hat{\mathbf{O}}(\boldsymbol{\Gamma}^{i,j,k},\hat{s},\hat{\bm{\mathsf{t}}},\hat{\boldsymbol{\partial}})}=\int\mathrm{d}^3\pos~\boldsymbol{\Omega}^{i,j,k}(\pos)
\end{equation}
With explicit expressions for quasi-relativistic ZORA density functions
given in eqs. \prettyref{eq: 1eMODF_split}, \prettyref{eq: twocompDF4c}
and \prettyref{eq: twocompAO4c}.
and the definitions of the relativistic density functions in eq.
\prettyref{eq: diracdensity} the working equation for the computation of
the expectation value of an arbitrary relativistic one-electron properties 
within ZORA is
\begin{widetext}
\begin{equation}
\Braket{\hat{\mathbf{O}}(\boldsymbol{\Gamma}^{i,j,k},\hat{s},\hat{\bm{\mathsf{t}}},\hat{\boldsymbol{\partial}})}=
\Sum{IJ \in \mathcal{M}(i)}{}
\Sum{n}{N_\mathrm{occ}}\Sum{\mu\nu}{N_\mathrm{basis}}
\Sum{l\in M(I)}{}
\Sum{m\in M(J)}{}
\mathcal{D}^{j,k,(l,m)}_{n\mu\nu}\circ\int\limits_{\pos'=\pos}\mathrm{d}^3\pos~
\brackets{\hat{s}(\pos)\hat{\bm{\mathsf{t}}}(\pos)\circ\hat{\boldsymbol{\partial}}(\pos'\vee\pos)\mathcal{I}^{(l,m)}_{\mu\nu}(\pos,\pos')}.
\label{eq: workingequation}
\end{equation}
\end{widetext}
Here $\mathcal{M}(i)$ maps the index $i$ of the Dirac density
function to a sum of two-component density functions with appropriate
sign following eqs. \prettyref{eq: diracdensity}.

The above expressions are valid for properties within first order perturbation theory
only. In case of second order properties the corresponding
density functions have to be constructed with perturbed
density matrices $\tilde{\mathbf{D}}^{(k)}$ or rather transition density
matrices. These can be received via common algorithms by solving the
coupled perturbed Hartree-Fock (CPHF) or Kohn-Sham (CPKS) equations.
For uncoupled problems the above density functions can be used in a
simple sum-over-states (SoS) framework. For this purpose, instead of
the MO-coefficients of only occupied orbitals, also the MO-coefficients
of unoccupied orbitals have to be considered, weighted by the orbital energy
differences.
\section{Implementation}
The above formalism was implemented within a two-component ZORA
version\cite{wullen:2010} of the quantum chemistry program package
Turbomole\cite{ahlrichs:1989}. The derivative operator
$\hat{\boldsymbol{\partial}}(\pos'\vee\pos)$ was limited to first and
second derivatives. The tensor operator can always be reduced to any
sum of tensor products of the 
electronic position vector with respect to some arbitrary origin
$\pos-\pos_\mathrm{O}$ or with respect to the center of a nucleus $A$:
$\pos-\pos_A$. Eq. \prettyref{eq:
workingequation} is rewritten in terms of two-component
density matrices $\mathbf{D}^{(k)}$:
\begin{widetext}
\begin{equation}
\Braket{\hat{\mathbf{O}}(\boldsymbol{\Gamma}^{i,j,k},\hat{s},\hat{\bm{\mathsf{t}}},\hat{\boldsymbol{\partial}})}=
\Sum{IJ \in \mathcal{M}(i)}{}
\Sum{n}{N_\mathrm{occ}}\Sum{\mu\nu}{N_\mathrm{basis}}\Sum{m=0}{3}
\mathfrak{Re}\braces{\imath^l D^{(m)}_{n\mu\nu}}\circ\underbrace{\int\limits_{\pos'=\pos}\mathrm{d}^3\pos~
\brackets{\hat{s}(\pos)\hat{\bm{\mathsf{t}}}(\pos)\circ\hat{\boldsymbol{\partial}}(\pos'\vee\pos)\tilde{\mathcal{I}}^{IJ,j,k,(m,l)}_{\mu\nu}(\pos,\pos')}
}_{O^{IJ,j,k,(m,l)}_{\mathrm{AO},\mu\nu}}
.
\label{eq: workingequation_true}
\end{equation}
\end{widetext}
Within above mentioned restrictions for the derivative operator, a
fortran code was automatically generated for evaluation of the
integrand of eq. \prettyref{eq: workingequation_true} (term in
square brackets) with the computer algebra system
Mathematica.\cite{mathematica10}

The code was generated for the derivative operators $\nabla\otimes$,
$\nabla\cdot$, $\nabla\times$, $\nabla\otimes\nabla\cdot$,
$\nabla\cdot\nabla\otimes$, $\nabla\otimes\nabla\times$, and
$\nabla\times\nabla\times$. For computation of maximally second 
derivatives, at most third derivatives of basis functions are required due to
appearance of first derivatives of matrix elements containing
the small component. Furthermore, up to second
derivatives of the model potential are needed with respect to $\pos$.
These were implemented following Ref. \onlinecite{wullen:1998}. In
case of second derivatives, third derivatives of density functionals
with respect to the model density are required.  These were
approximated by central finite differences of analytical second
derivatives of the density functional. 

Within our implementation, integrals are evaluated by default on a grid
using standard numerical integration methods, established for density
functional theory (DFT) calculations. Integrals that do not contain the
ZORA-factor $\omega$ can typically be evaluated analytically. In the
present implementation, however, only a few analytical integrals are
available.  

We added also a SoS module for calculation of second order properties
within an uncoupled DFT approach. The SoS calculation is carried out
with transition density functions which are evaluated analogously to
eqs. \prettyref{eq: 1eMODF1} to \prettyref{eq: 1eMODF4} as densities between occupied
and unoccupied orbitals $i,a$:
\begin{equation}
\tilde{\rho}^{\mathrm{LL},j,k}_{ia}(\pos,\pos')=
\mathfrak{Re}
\Sum{\mu\nu}{N_\text{basis}}
\braces{\vec{C}_{i\mu}^\dagger\boldsymbol{\varsigma}(j,k)\vec{C}_{a\nu}}
~\cdot~
\parantheses{\chi_\mu\chi'_\nu}
\end{equation}
and so on.  The second order SoS expression is
\begin{multline}
\Sum{i=1}{N_\text{occ}}
\Sum{a=1+N_\text{occ}}{N_\text{orb}}
\frac{
\Braket{\psi_i|\Op{\mathbf{O}}_1|\psi_a}
\Braket{\psi_a|\Op{\mathbf{O}}_{2}|\psi_i}
}{
\parantheses{\epsilon_i-\epsilon_{a}}
} +~\text{cc},
\label{eq: sos}
\end{multline}
where $\epsilon$ are the orbital energies, $i$ are indices of occupied
and $a$ are indices of unoccupied orbitals. The integrals
$\Braket{\psi_i|\Op{\mathbf{O}}|\psi_a}$ are evaluated via the corresponding
generic transition one-electron tensor functions:
\begin{equation}
\Braket{\psi_i|\Op{\mathbf{O}}|\psi_a}=\int\mathrm{d}^3\pos~\tilde{\boldsymbol{\Omega}}^{j,k}_{ia}(\pos),
\end{equation}
with
\begin{equation}
\tilde{\boldsymbol{\Omega}}^{IJ,j,k}_{ia}(\pos)=
\hat{s}\hat{\bm{\mathsf{t}}}\circ\left.\hat{\boldsymbol{\partial}}(\pos'\vee\pos)
\tilde{\rho}^{IJ,j,k}_{ia}(\pos,\pos')\right|_{\pos'=\pos}.
\end{equation}
The contraction of AO-matrix elements with LCAO-coefficients is
implemented via density matrices as described in the last section or
matrix multiplications of the type
$(\mathbf{C}_\mathrm{occ}^{(\xi)})^{\dagger}\mathbf{O}^{IJ,j,k,(m,l)}_\mathrm{AO}\mathbf{C}_\mathrm{unocc}^{(\xi)}$,
where $\xi$ can be $\alpha$ or $\beta$,
$\mathbf{C}_\mathrm{occ}^{(\xi)}$ and
$\mathbf{C}_\mathrm{unocc}^{(\xi)}$ are the $\alpha$- and $\beta$-spin
blocks of the block of occupied and unoccupied orbitals in the
coefficient matrix, respectively.
Here, for component $(m,l)$ of the matrix of integrals in AO-basis
$\mathbf{O}^{IJ,j,k,(m,l)}_\mathrm{AO}$ the sum of matrix
multiplications with $\alpha$- and $\beta$-coefficient matrices, that
correspond to $\boldsymbol{\varsigma}(m,l)$ is formed.  

In order to account for renormalisation effects, which are in the ZORA
approach particularly relevant for the description of the energetically
lowest lying orbitals with main contributions close to the nucleus, we
implemented the possibility of renormalisation of the ZORA wave function by
redefinition of the coefficient matrices:
\begin{equation}
\vec{\tilde{C}}_{i\mu} =
\frac{\vec{C}_{i\mu}}{\sqrt{1+\int\mathrm{d}^3\pos~\rho^{\mathrm{SS},0,0}_i(\pos)}}.
\label{eq: renorm}
\end{equation}
This renormalisation can be evaluated directly within the
general implementation of one-electron operators presented above.
\section{Computational details}
Quasi-relativistic two-component calculations are performed within 
ZORA at the level of complex generalized Hartree--Fock (cGHF) or
Kohn--Sham (cGKS) with a modified 
version\cite{wullen:2010} of the
quantum chemistry program package Turbomole\cite{ahlrichs:1989}. 

For Kohn--Sham (KS)-DFT calculations the hybrid Becke three parameter exchange
functional and Lee, Yang and Parr correlation functional
(B3LYP)\cite{stephens:1994,vosko:1980,becke:1988,lee:1988} was employed.
In comparison to relativistic coupled cluster (CC) calculations this
functional performed well for the description of $\mathcal{P,T}$-odd
effects in diatomic radicals in our previous
work, which motivates the present
choice.\cite{gaul:2017,gaul:2019,gaul:2018a}

For all calculations a basis set of 37~s, 34~p, 14~d and 9~f
uncontracted Gaussian functions with the exponential coefficients
$\alpha_i$ composed as an even-tempered series by $\alpha_i=a\cdot
b^{N-i};~ i=1,\dots,N$, with $b=2$ for s- and p-function and with
$b=(5/2)^{1/25}\times10^{2/5}\approx 2.6$ for d- and f-functions was
used for Tl and Ra. The largest exponent coefficients of the s, p, d and f 
subsets are $2\times10^9~a_0^{-2}$, $5\times10^8~a_0^{-2}$,
$13300.758~a_0^{-2}$
and $751.8368350~a_0^{-2}$,
respectively. This basis set has proven successful in
calculations of nuclear-spin dependent $\mathcal{P}$-violating
interactions and $\mathcal{P,T}$-odd effects induced by an eEDM in
heavy polar diatomic
molecules.\cite{isaev:2012,Isaev:2013,isaev:2014,gaul:2017,gaul:2019,gaul:2018a}
The F atom was represented with a decontracted atomic natural orbital
(ANO) basis set of triple-$\zeta$ quality\cite{roos:2004}. 

The ZORA-model potential $\tilde{V}(\pos)$ as proposed by van
W\"ullen\cite{wullen:1998} was employed with
additional damping\cite{liu:2002}.

For calculations of two-component wave functions and properties a
finite nucleus was used, described by a normalized spherical Gaussian
nuclear density distribution
$\rho_{\mathrm{nuc},A}(\pos)=-\frac{\zeta_A^{3/2}}{\pi^{3/2}}\mathrm{e}^{-\zeta_A\abs{\pos-\pos_A}^2}$,
where $\zeta_A= \frac{3}{2r_{\text{nuc},A}^2}$ and the root mean
square radius $r_{\text{nuc},A}$ of nucleus $A$ was used as suggested
by Visscher and Dyall.\cite{visscher:1997} The mass numbers $A$ were chosen
to correspond to the isotopes $^{205}$Tl,$^{223}$Ra. 

Nuclear equilibrium distances were obtained at the levels of GHF-ZORA and
GKS-ZORA/B3LYP, respectively. As convergence criteria an energy change
of less than $10^{-5}~E_\text{h}$ was used. For DFT calculations of analytic energy
gradients with respect to the displacement of the nuclei the nuclei
were approximated as point charges. The equilibrium 
distances obtained are for $^{205}$TlF 2.08~\AA{} (cGHF) and 2.12~\AA{}
(cGKS) and for $^{223}$RaF 2.28~\AA{} (cGHF) and 2.26~\AA{} (cGKS).

All properties were computed with and without renormalisation of the
wave function according to \prettyref{eq: renorm}.
\section{Results and discussion}
\subsection{Enhancement of various sources of $\mathcal{P,T}$-violation
in paramagnetic RaF and diamagnetic TlF}
A measurement of a permanent electric dipole moment in vanishing
electric field would indicate a simultaneous violation of parity
$\mathcal{P}$ and time-reversal $\mathcal{T}$ symmetry (see e.g
Ref. \onlinecite{khriplovich:1997}). Molecular
systems provide currently the strictest experimental limits on permanent 
electric dipole moments.\cite{demille:2015,andreev:2018} 

There are many possible $\mathcal{P,T}$-violating sources that can
lead to a permanent electric dipole moment in a molecule depending on
the nuclear and electronic spin states of the
molecule.\cite{khriplovich:1997,kozlov:1995} For
paramagnetic molecules such as RaF pronounced
sensitivity is expected for a potential permanent electric dipole moment
of the electron (eEDM) $d_\mathrm{e}$ and from $\mathcal{P,T}$-odd scalar-pseudoscalar
nucleon-electron current (SPNEC) interactions
$k_\mathrm{S}$.\cite{ginges:2004} 

However, in case of a non-zero nuclear spin additional contributions
from interactions of the electron cloud with a potential permanent
electric dipole moment of the proton (pEDM) $d_\mathrm{p}$,
interactions with a net electric dipole moment of the nucleus, called
Schiff moment $\mathcal{S}$, tensor-pseudotensor nucleon-electron
current (TPNEC) interactions $k_\mathrm{T}$, and pseudoscalar-scalar
nucleon-electron current (PSNEC) interactions $k_\mathrm{p}$ can
occur.  The latter, PSNEC, vanishes in the limit of infinitely
large mass of the nucleus, which appears to be a reasonable approximation for 
heavy-elemental molecules, as 
$m_\mathrm{e}\ll m_\mathrm{nuc}$.\cite{khriplovich:1997}

Nuclear spin-dependent contributions are the dominating sources for
$\mathcal{P,T}$-violation in diamagnetic molecules such as TlF, where
interactions with an eEDM or due to SPNECs appear only as indirect
interactions via hyperfine induced coupling, because of vanishing total
effective electron spin.

In paramagnetic molecules containing nuclei with spin quantum number $I$
larger than $^1/_2$, interactions with potential higher $\mathcal{P,T}$-odd
nuclear moments, as e.g. nuclear magnetic quadrupole moments (NMQM), could
contribute to a permanent molecular EDM.

The full $\mathcal{P,T}$-odd effective spin-rotational Hamiltonian for a
paramagnetic diatomic molecule for nucleus $A$ with nuclear spin quantum
number larger than $^1/_2$ reads (for parts of
${\Op{H}_{\mathrm{sr}_{A}}}$ see Refs. \onlinecite{kozlov:1995} and
\onlinecite{hinds:1980})
\begin{widetext}
\begin{equation}
\begin{aligned}[t]
{\Op{H}_{\mathrm{sr}_{A}}} &=
\underbrace{\vec{\lambda}\cdot\Op{\vec{S}}}_{\Omega}
\parantheses{W_{\mathrm{d}_A}d_\mathrm{e}+W_{\mathrm{s}_A}
k_\mathrm{s}}
+\underbrace{\vec{\lambda}^T\cdot\Op{\mathbf{T}}_A\cdot\Op{\vec{S}}}_{\Theta_A}
W_{\mathcal{M}_A}\tilde{\mathcal{M}}_A + \text{higher moments\dots}\\
&+\underbrace{\vec{\lambda}\cdot\Op{\vec{I}}_A}_{\mathcal{I}_A}
\parantheses{
W_{\mathrm{T}_A} k_\mathrm{T}
+W_{\mathrm{p}_A} k_\mathrm{p}
+W^\mathrm{m}_{\mathrm{s}_A} k_\mathrm{s}
+W_{\mathcal{S}_A} \mathcal{S}_A
+(W_{\mathrm{m}_A}+W_{\mathcal{S}_A}R_\text{vol}) d_\mathrm{p}
+W^\mathrm{m}_{\mathrm{d}_A} d_\mathrm{e}
}.
\end{aligned}
\end{equation}
\end{widetext}
Here $\vec{\lambda}$ is the unit vector pointing from the
heavy to the light nucleus, $\hat{\vec{S}}$ is the effective electron spin of the
molecule, $\hat{\vec{I}}_A$ the effective nuclear spin of nucleus $A$,
$\Op{\mathbf{T}}_A$ is a second-rank tensor which can be constructed
from the components of $\hat{\vec{I}}_A$ (for details see Ref.
\onlinecite{kozlov:1995}),
$\tilde{\mathcal{M}}_A=\frac{-1}{2I_A(2I_A-1)}\mathcal{M}_A$ with the NMQM
$\mathcal{M}$ (see Ref. \onlinecite{kozlov:1995}),
$\Omega$ is the projection of the effective electron spin on the
molecular axis and $\mathcal{I}_A$ is the projection of
$\hat{\vec{I}}_A$ on the molecular axis. For a diamagnetic molecule
all terms that depend on the effective electronic spin $\vec{S}$ (here
all proportional to $\Omega$ and $\Theta$) vanish. The constants $W$ are
electronic structure coupling constants enhancing
$\mathcal{P,T}$-violating parameters in molecules, that need to be
determined by electronic structure calculations.  $R_\text{vol}$ is a
nuclear structure factor that enhances the pEDM and can be determined
from nuclear structure calculations.

In the following we will focus on the electronic structure enhancement
factors $W$ in RaF and TlF. Thereby we will not calculate
$W_{\mathrm{p}_A}$ which is supposed to be many orders of magnitude smaller
than the other effects.\cite{khriplovich:1997,ginges:2004} Furthermore, we
do not include in our discussion those nuclear-spin dependent effects
$W^\mathrm{m}_{\mathrm{s}_A}$ and $W^\mathrm{m}_{\mathrm{d}_A}$ that are
induced by hyperfine coupling and thus are second-order molecular
properties that can be obtained from a linear response treatment. Assuming
the molecular axes to be aligned along the $z$-axis the remaining
electronic structure parameters are defined in the following (see e.g
\onlinecite{khriplovich:1997,kozlov:1995,martensson-pendrill:1987,hinds:1980,quiney:1998a}):

i) electronic structure enhancement of the eEDM 
\begin{equation}
W_{\mathrm{d}}=\frac{\Braket{\Psi|\frac{2c}{e\hbar}\imath\diraccontra{0}\diraccontra{5}\momop^2|\Psi}}{\Omega},
\end{equation}
with $e$ being the elementary charge and $\hbar=\frac{h}{2\pi}$ being
the reduced Planck's constants;
 
ii) electronic structure enhancement of the NMQM 
\begin{equation}
W_{\mathcal{M}_A}=\frac{\Braket{\Psi|cek_\mathrm{em}\frac{3}{2}\frac{(z-z_A)\parantheses{\diraca\times(\pos-\pos_A)}_z}{\abs{\pos-\pos_A}^5}
|\Psi}}{\Omega},
\end{equation}
with the constant $k_\mathrm{em}$ being $\frac{\mu_0}{4\pi}$ in SI
units with $\mu_0$ being the magnetic constant (see Ref.~\onlinecite{cohen:2008} for
other choices of $k_\mathrm{em}$ that correspond to different unit systems);
 
iii) electronic structure enhancement of nuclear Schiff moment and
volume effect due to a pEDM
\begin{equation}
W_{\mathcal{S},A}=\left.\frac{2\pi}{3} \partiell{}{z}
\Gamma^{0,0,0}(\pos,\pos)\right|_{\pos=\pos_A},
\end{equation}
in calculations of $W_{\mathcal{S},A}$ the value was calculated at
eight points at a distance of $1.7\times10^{-25}~a_0$ around the
nucleus (cubic arrangement) and averaged. This Hamiltonian is a
consequence of Schiff's theorem\cite{schiff:1963} and is the dipole 
contribution of an expansion of the electric potential of a finite
nucleus (for details see Ref. \onlinecite{sushkov:1984});

iv) electronic structure enhancement of the pEDM due to magnetic
fields of moving electrons
\begin{equation}
W_{\mathrm{m}_A}=\Braket{\Psi|4\parantheses{\frac{\mu_\mathrm{N}}{A_A 
}+\frac{\mu_A}{Z_A}}\frac{c}{\hbar}\frac{k_\mathrm{em}}{k}\frac{\parantheses{\diraca\times\hat{\vec{\ell}}_A}_z}{\abs{\pos_A}^3}|\Psi},
\end{equation}
with constant $k$, which is 1 in SI units and $c^{-1}$ in Gauss units, the
orbital angular momentum operator with respect to nucleus $A$
$\hat{\vec{\ell}}_A=(\pos-\pos_A)\times\momop$, the nuclear magneton
$\mu_\mathrm{N}=\frac{e\hbar}{2m_\mathrm{p}}$ with the mass of the proton
$m_\mathrm{p}$, the nuclear magnetic moment $\mu_A$,
the nuclear charge $Z_A$ and mass number $A_A$ of nucleus $A$. Here we
assume that the contributions stem from a single active valence proton
in the nuclear shell (see Ref. \onlinecite{hinds:1980});

v) electronic structure enhancement of SPNEC interactions
\begin{equation}
W_{\mathrm{s}_A}=\frac{\Braket{\Psi|\frac{G_\mathrm{F}}{\sqrt{2}}\imath\diraccontra{0}\diraccontra{5}\rho_{\mathrm{nuc},A}|\Psi}}{\Omega},
\end{equation}
with the Fermi weak coupling constant
$G_\mathrm{F}=2.22249\times10^{-14}E_\mathrm{h}a_0^3$;
 
vi) electronic structure enhancement of TPNEC interactions
\begin{equation}
W_{\mathrm{T}_A}=\Braket{\Psi|\sqrt{2}G_\mathrm{F}\imath\diraccontra{3}\rho_{\mathrm{nuc},A}|\Psi}\,.
\end{equation}

With the method presented in the previous section all these properties
can be directly evaluated as illustrated in \prettyref{tab:
howtousegenden}. We choose as exemplary molecular systems RaF and TlF,
which were well studied before and are considered promising candidates
for experiments that aim at a measurement of $\mathcal{P,T}$-odd
properties.\cite{Isaev:2013,kudashov:2014,sasmal:2016a,hinds:1980,parpia:1997,quiney:1998a} 

We neglect any magnetic or many-electron effects on the above
presented properties that may arise from the ZORA transformation as
these are expected to be low for heavy elements. For some of the
properties such effects were discussed elsewhere (see e.g. Refs.
\onlinecite{lindroth:1989,gaul:2019}).

\subsubsection{Diamagnetic Molecules: TlF}

In this section we discuss results for TlF that were obtained from
property calculations with an implementation of the approach described
in the previous sections.  Results of our calculations on the level of
cGHF- and cGKS-B3LYP-ZORA are compared to results from literature in
\prettyref{tab: tlfresults}.

We see an excellent agreement (deviations $\leq5~\%$) between
cGHF-ZORA and DHF results reported by Quiney \emph{et. al.} for all
calculated properties.\cite{quiney:1998a}  Furthermore,
renormalisation of the wave function does not play an important role
for the nuclear spin-dependent properties and is always below 1~\%.
This is to be expected as all major contributions stem from the
valence molecular orbitals at the position of the nucleus. 
This shows that ZORA is appropriate for the quantitative description
of relativistic effects due to nuclear-spin dependent
$\mathcal{P,T}$-odd interactions in a heavy molecule, which reinforces
related previous findings for nuclear-spin independent and
nuclear-spin dependent $\mathcal{P}$-odd interactions in molecules
\cite{berger:2005,berger:2005a,berger:2007,nahrwold:09,isaev:2010,isaev:2012,isaev:2014}.
Correlation effects were estimated on the DFT level with the B3LYP
hybrid density functional. This functional, however, seems to
overshoot electron correlation effects leading to values being too low
in magnitude with deviations of up to $23~\%$ in comparison to
GRECP/RCC-SD calculations.\cite{petrov:2002}

\subsubsection{Paramagnetic Molecules: RaF}
We report results on electronic structure enhancement
factors of nuclear spin-dependent and nuclear spin-independent $\mathcal{P,T}$-effects
in $^{223}$RaF in \prettyref{tab: rafresults} and compare to all
available literature data.

Also for nuclear-spin independent properties and the nuclear- and
electron-spin dependent NMQM, most important contributions stem from
valence molecular orbitals and thus effects of renormalisation are
negligible.
The reasonable agreement between cGHF/cGKS-ZORA and four component
coupled cluster calculations\cite{kudashov:2014,sasmal:2016a} for
$W_\mathrm{d}$ and $W_\mathrm{s}$ was discussed
elsewhere.\cite{kudashov:2014,gaul:2017,gaul:2019} Our present values for
$W_\mathrm{d}$ and $W_\mathrm{s}$ 
differ in the last reported digit from the results in Refs.
\onlinecite{gaul:2017,gaul:2019} as we consider herein a different isotope
of Ra. Our calculations of
$W_\mathcal{S}$ are with deviations of $8~\%$ (cGHF) in good
agreement with GRECP-FSCC calculations (Ref.
\onlinecite{kudashov:2014}). 
As observed in previous
studies,\cite{gaul:2017,gaul:2019} DFT tends to give results too low
in absolute value. However, the value of $W_\mathcal{S}$ still agrees
reasonably (deviation is $12~\%$) with the coupled cluster
calculations. From this we expect a similar precision of cGHF- and
cGKS-ZORA calculations of $W_\mathrm{T}$ and $W_\mathrm{m}$.

Our calculations show that nuclear-spin dependent
$\mathcal{P,T}$-odd enhancement factors in RaF are by about a factor
of $1/2$ to $1/8$ smaller in magnitude compared to TlF. The
ratios of $\frac{W_\mathrm{T}}{W_\mathcal{S}}$ and
$\frac{W_\mathrm{m}}{W_\mathcal{S}}$ are considerably different for
TlF and RaF. Thus data from measurements of RaF would complement data
from TlF measurements as different regions in the parameter space of
the $\mathcal{P,T}$-odd parameters $k_\mathrm{T}$, $d_\mathrm{p}$ and
$\mathcal{S}$ are covered.

In \prettyref{tab: rafresults} we presented values for
$W_\mathcal{M}$, as well. In comparison to predictions of NMQM
enhancement made for other molecules\cite{flambaum:2014} the values
for RaF are very large (as large as predicted for YbF
or ThO). Furthermore the $^{223}$Ra nucleus is known to have a
octupole deformation, which is expected to enhance NMQM effects
significantly on the nuclear structure level.\cite{flambaum:1994} This
makes $^{223}$RaF a promising candidate for setting strict limits on NMQM
induced permanent electric dipole moments.

\section{Conclusion}
We outlined a generally applicable approach to the evaluation of
arbitrary relativistic properties within an approximate
quasi-relativistic wave function. Automated code generation via a
computer algebra system was applied to obtain a pilot implementation within
a modified version of a quantum chemical program for
ZORA wave functions. Within the approach presented herein, relativistic
first and second order properties ranging from commonly available molecular
properties, such as NMR-shielding constants, to less common discrete
symmetry violating properties are accessible in a single
implementation. The flexibility of this property toolbox was demonstrated
by computation of a number of $\mathcal{P,T}$-odd effects that are
important for fundamental physics research with diatomic molecules

Within this study large enhancements of the nuclear magnetic
quadrupole moment and nuclear spin-dependent as well as nuclear
spin-independent sources of $\mathcal{P,T}$-violation in RaF were
determined. This shows that $^{223}$RaF is a well suited system for
setting strict limits on $\mathcal{CP}$-violation in essentially all sectors of
particle physics. 

\begin{acknowledgments}
Computer time provided by the center for scientific computing (CSC)
Frankfurt and financial support by the Deutsche Forschungsgemeinschaft
via Sonderforschungsbereich 1319 (ELCH) ``Extreme Light for
Sensing and Driving Molecular Chirality'' are gratefully
acknowledged.
\end{acknowledgments}

%
\clearpage

\begin{table*}
\begin{threeparttable}
\caption{Representation of various $\mathcal{P,T}$-odd properties in
the formalism of general tensor functions.}
\label{tab: howtousegenden}
\begin{tabular}{cccccc}
\toprule
Property                            & 
Prefactor                           &
$\hat{s}$                           & 
$\hat{\bm{\mathsf{t}}}\circ$                &
$\hat{\boldsymbol{\partial}}\circ$  & 
$\boldsymbol{\Gamma}^{i,j,k}$ \\ 
\midrule
{$W_\mathrm{d}$}                  &$ \frac{2c\hbar}{e\Omega} $
                                  & - 
                                  & - 
                                  &$ \nabla\cdot\nabla\otimes $
                                  &$ \boldsymbol{\Gamma}^{2,0,1} $\\
{$W_\mathcal{M}$}                 &$
\frac{3}{2\Omega}cek_\mathrm{em} $
                                  &$ \frac{z-z_A}{\abs{\pos-\pos_A}^5} $
                                  &$ -(\pos-\pos_A)\times $
                                  & -
                                  &$\vec{\boldsymbol{\Gamma}}^{1,(1,2,3),0} $\\
{$W_{\mathcal{S}}$}&$ \frac{2\pi}{3} $
                                  &$ \delta(\pos_A-\pos) $
                                  & -  
                                  &$ \nabla\otimes $ 
                                  &$ \boldsymbol{\Gamma}^{0,0,0} $\\
{$W_\mathrm{m}$}                  &$4\parantheses{\frac{\mu_\mathrm{N}}{A_A 
}+\frac{\mu_A}{Z_A}}\frac{c}{\hbar}\frac{k_\mathrm{em}}{k}  $
                                  &$ \abs{\pos-\pos_A}^{-3} $
                                  &$ (\pos-\pos_A)\times $
                                  &$ \nabla\times $
                                  &$ \vec{\boldsymbol{\Gamma}}^{1,(1,2,3),1} $\\
{$W_\mathrm{s}$}                  &$ \frac{G_\mathrm{F}}{\sqrt{2}\Omega} $
                                  &$ \rho_\mathrm{nuc}(\pos) $
                                  & -  
                                  & -
                                  &$ \boldsymbol{\Gamma}^{2,0,1}  $\\
{$W_\mathrm{T}$}                  &$ \sqrt{2}G_\mathrm{F} $
                                  &$ \rho_\mathrm{nuc}(\pos)  $
                                  & -
                                  & -  
                                  &$ \boldsymbol{\Gamma}^{2,3,1} $\\
\bottomrule
\end{tabular}
\end{threeparttable}
\end{table*}
\begin{table*}
\begin{threeparttable}
\caption{Nuclear spin-dependent, electron spin-independent
$\mathcal{P,T}$-odd electronic structure parameters of the diamagnetic
molecule $^{205}$TlF evaluated at the level of cGHF- and cGKS-ZORA with a
large even tempered basis set with (wr) and without renormalisation (wor) of
the density according to eq. 
\prettyref{eq: renorm}. Comparison to literature values
determined with different computational method. The value
$\mu(^{205}\mathrm{Tl})=1.6382135~\mu_\mathrm{N}$ was used for
the nuclear magnetic moment of $^{205}$Tl.\cite{mills:1993}}
\label{tab: tlfresults}
\begin{tabular}{l
S[table-number-alignment=center,table-figures-decimal=0]
S[table-number-alignment=center,table-figures-decimal=2]
S[table-number-alignment=center,table-figures-decimal=0]
}
\toprule
Method 
&{$W_\mathrm{T}/h\,\mathrm{Hz}$}
&{$W_\mathrm{m}/\frac{10^{18}~h\,\mathrm{Hz}}{e\,\mathrm{cm}}$}
&{$W_{\mathcal{S}}/a_0^4$}\\
\midrule
cGHF-ZORA-wor                                       &4697&-4.74&8443\\
cGHF-ZORA-wr                                        &4690&-4.72&8428\\
cGKS-ZORA-B3LYP-wr                                  &3375&-3.10&5720\\
DHF\tnote{a} (Ref. \onlinecite{quiney:1998a})       &4632&-4.78&8747\\
GRECP-RCC-SD\tnote{b} (Ref. \onlinecite{petrov:2002})&{-}&-4.04&7635\\
DF\tnote{c} (Ref. \onlinecite{parpia:1997})          &{-}&-5.46&7738\\
\bottomrule
\end{tabular}
\begin{tablenotes}\footnotesize 
\item[a]Dirac--Hartree--Fock calculation without electron correlation.

\item[b] Generalized relativistic effective core potential, two-step
approach with restricted active space SCF electron-correlation
calculation at the level of single and double excitations.

\item[c]Dirac-Fock calculation without electron-correlation.
\end{tablenotes}
\end{threeparttable}
\end{table*}
\begin{table*}
\begin{threeparttable}
\caption{Electron and nuclear spin-dependent $\mathcal{P,T}$-odd
electronic structure parameters of the paramagnetic molecule
$^{223}$RaF evaluated at the level of cGHF- and cGKS-ZORA with a large
even tempered basis set with (wr) and without renormalisation (wor) of
the density according to eq. 
\prettyref{eq: renorm}.  Comparison to available literature values
determined with different computational methods. The value
$\mu(^{223}\mathrm{Ra})=0.271~\mu_\mathrm{N}$ was used for
the nuclear magnetic moment of $^{223}$Ra.\cite{mills:1993} In all our
calculations $\Omega=0.500$.}
\label{tab: rafresults}
\begin{tabular}{l
S[table-number-alignment=center,table-figures-decimal=0]
S[table-number-alignment=center,table-figures-decimal=2]
S[table-number-alignment=center,table-figures-decimal=0]
S[table-number-alignment=center,table-figures-decimal=0]
S[table-number-alignment=center,table-figures-decimal=2]
S[table-number-alignment=center,table-figures-decimal=2]
}
\toprule
Method 
&{$W_\mathrm{T}/h\mathrm{Hz}$}
&{$W_\mathrm{m}/\frac{10^{18}~h\,\mathrm{Hz}}{e\,\mathrm{cm}}$}
&{$W_\mathcal{S}/a_0^4$}
&{$W_\mathrm{s}/h\,\mathrm{kHz}$}
&{$W_\mathrm{d}/\frac{10^{24}~h\,\mathrm{Hz}}{e\,\mathrm{cm}}$}
&{$W_\mathcal{M}/\frac{10^{33}~h\,\mathrm{Hz}}{c\,e\,\mathrm{cm}^2}$}
\\
\midrule
cGHF-ZORA-wor                                &-1810&0.66&-4235
&-152&-27.2&-1.17\\
cGHF-ZORA-wr                                 &-1809&0.66&-4229
&-152&-27.2&-1.17\\
cGKS-ZORA-B3LYP-wr                           &-1617&0.58&-3686
&-138&-24.6&-1.03\\
FS-RCCSD+$\Delta_\text{basis}$+$\Delta_\text{triples}$\tnote{a}
(Ref.\onlinecite{kudashov:2014})   &{-}  &{-} &-4260&-139&-25.6&{-}\\
DF-CCSD\tnote{b} (Ref. \onlinecite{sasmal:2016a})&{-}  &{-} &{-}      &-141&-25.4&{-}\\
\bottomrule
\end{tabular}
\begin{tablenotes}\footnotesize 
\item[a] Relativistic two-component Fock-space coupled-cluster
approach with single and double excitations (CCSD) with basis set
corrections from CCSD calculations with normal and large sized basis
sets and triple excitation corrections from CCSD calculations with and
without perturbative triples.  

\item[b]Dirac-Fock calculation with electron-correlation effects on
the level of coupled cluster with single and double excitations.
\end{tablenotes}
\end{threeparttable}
\end{table*}
\end{document}